\documentstyle[12pt,amssymbols]{article}
\def\ltap{\raisebox{-.4ex}{\rlap{$\sim$}} \raisebox{.4ex}{$<$}}
\def\gtap{\raisebox{-.4ex}{\rlap{$\sim$}} \raisebox{.4ex}{$>$}}
\def\journal{\topmargin .3in    \oddsidemargin .5in
        \headheight 0pt \headsep 0pt
        \textwidth 5.625in % 1.2 preprint size  %6.5in
\textheight 8.25in % 1.2 preprint size 9in
        \marginparwidth 1.5in
        \parindent 2em
        \parskip .5ex plus .1ex         \jot = 1.5ex}
\journal

\newskip\humongous \humongous=0pt plus 1000pt minus 1000pt

\newif\ifdtup

\input{epsf}
\begin{document}
\begin{titlepage}
\begin{center}
%\date               \hfill   LBL-    \\
December 14, 1995   \hfill    LBL- 38044\\
%\hfill hep-ph/9503458
\vskip .5in

{\large \bf Strong $WW$ scattering in unitary gauge}
\footnote
{This work was supported by the Director, Office of Energy
Research, Office of High Energy and Nuclear Physics, Division of High
Energy Physics of the U.S. Department of Energy under Contract
DE-AC03-76SF00098.}

\vskip .5in

Michael S. Chanowitz\footnote{Email: chanowitz@lbl.gov}

\vskip .2in

{\em Theoretical Physics Group\\
     Lawrence Berkeley Laboratory\\
     University of California\\
     Berkeley, California 94720}
\end{center}

\vskip .25in

\begin{abstract}
A method to embed models of strong 
$WW$ scattering in unitary gauge amplitudes is presented 
that eliminates the need for the effective $W$ approximation (EWA) in
the computation of cross sections at high energy colliders.
The cross sections obtained from the U-gauge amplitudes 
include the distributions of the final state
fermions in $ff \rightarrow ffWW$, which cannot be obtained from the EWA. 
Since the U-gauge method preserves the interference of the signal and the
gauge sector background amplitudes, 
which is neglected in the EWA, it is more accurate,
especially if the latter is comparable to or bigger than the signal, 
as occurs for instance at small angles because of Coulomb
singularities. The method is illustrated for on-shell $W^+W^+ \rightarrow 
W^+W^+$ scattering and for $qq \rightarrow qqW^+W^+$.
\end{abstract}

\vskip .2in

\end{titlepage}

%THIS PAGE (PAGE ii) CONTAINS THE LBL DISCLAIMER
%TEXT SHOULD BEGIN ON NEXT PAGE (PAGE 1)
\renewcommand{\thepage}{\roman{page}}
\setcounter{page}{2}
\mbox{ }

\vskip 1in

\begin{center}
{\bf Disclaimer}
\end{center}

\vskip .2in

\begin{scriptsize}
\begin{quotation}
This document was prepared as an account of work sponsored by the United
States Government. While this document is believed to contain correct
 information, neither the United States Government nor any agency
thereof, nor The Regents of the University of California, nor any of their
employees, makes any warranty, express or implied, or assumes any legal
liability or responsibility for the accuracy, completeness, or usefulness
of any information, apparatus, product, or process disclosed, or represents
that its use would not infringe privately owned rights.  Reference herein
to any specific commercial products process, or service by its trade name,
trademark, manufacturer, or otherwise, does not necessarily constitute or
imply its endorsement, recommendation, or favoring by the United States
Government or any agency thereof, or The Regents of the University of
California.  The views and opinions of authors expressed herein do not
necessarily state or reflect those of the United States Government or any
agency thereof, or The Regents of the University of California.
\end{quotation}
\end{scriptsize}

\vskip 2in

\begin{center}
\begin{small}
{\it Lawrence Berkeley Laboratory is an equal opportunity employer.}
\end{small}
\end{center}

\newpage

\renewcommand{\thepage}{\arabic{page}}
\setcounter{page}{1}
%THIS IS PAGE 1 (INSERT TEXT OF REPORT HERE)
%starthere

\noindent {\it \underline {Introduction} }

Electroweak symmetry breaking may be 
due to a weak or strong force. In the first case there are Higgs
bosons lighter than 1 TeV. In the second case there is
strong scattering of longitudinally polarized $W$ bosons at energies
$\sqrt{s_{WW}}\ \gtap\ 1$ TeV.
By measuring $WW$ scattering in the process $qq \rightarrow qqWW$ 
at a high energy collider such as the LHC
we will determine definitively which choice nature has made. 

The strong $WW$ scattering 
cross sections at  high energy colliders have traditionally been estimated by
combining the use of the 
equivalence theorem (ET) and the effective W approximation 
(EWA).\cite{mcmkg2}   The ET\cite{etrefs} 
represents strong $W_L W_L$ scattering ($L$ denotes longitudinal
polarization) in terms of the corresponding R-gauge, unphysical 
Goldstone boson $ww$ scattering amplitude, 
$$
{\cal{M}}(W_L(p_1), W_L(p_2), \ldots ) = {\cal{M}} (w(p_1), w(p_2),\ldots )_R
+{\rm O}\left(g^2,{M_W\over E_i}\right).\eqno(1)
$$
($g$ is the weak $SU(2)_L$ coupling constant)
so that a model of Goldstone boson scattering becomes a model for strong 
gauge boson scattering at high enough energy. 
Convoluting $\sigma(W_L W_L \rightarrow W_L W_L)$ 
obtained from the ET with the effective $W_L W_L$ luminosity from the 
EWA\cite{ewarefs} 
$$
\left. {d{\cal L}\over dz}\right |_{W_LW_L/qq} = {\alpha_W^2 \over 16\pi^2} 
{1 \over z} \left[ (1+z){\rm ln}{1 \over z} -2 + 2z \right]
,\eqno(2)
$$
where $z=s_{WW}/s_{qq}$, the parton subprocess cross section from the ET-EWA
method is 
$$
\sigma(qq \rightarrow qqW_LW_L) = \int dz {d{\cal L}\over dz} 
\sigma(W_LW_L \rightarrow W_LW_L).\eqno(3)
$$
The total cross section $\sigma(qq \rightarrow qqWW)$ is obtained by
incoherently adding the signal and background cross sections, with the latter
obtained from the standard model with a light or massless Higgs boson, say 
$m_H \ltap 100$ GeV.

The EWA is a good approximation for strong $W_LW_L$ scattering\footnote{
For several reasons the EWA is typically less useful for scattering of 
transverse modes.}
within its domain of applicability, defined by 
energies $E \gg M_W$ and scattering angles big enough 
that Coulomb singularities from photon exchange are not too large.
However because the EWA is obtained from a small angle approximation 
the transverse momenta of the final state jets (and the $WW$
diboson) are neglected and the transverse momentum distributions of the 
individual $W$ bosons are distorted.
Furthermore, because the EWA neglects the
interference between symmetry breaking sector (signal) and gauge 
sector (background) amplitudes, it may fail if the
signal is not much bigger than the background, as occurs for instance 
near Coulomb singularities.
These problems are both addressed by a
method presented here in which strong scattering models, formulated as
usual in an R-gauge by means of the ET, 
are ``transcribed'' to the complete set of U-gauge tree
amplitudes, for $WW \rightarrow WW$ or $ff \rightarrow ffWW$. 
Collider cross sections are then obtained directly from the 
$ff \rightarrow ffWW$ amplitudes without resorting to the EWA.
The momentum distributions of the final state quanta and the 
interference terms are automatically retained.

In discussing strong scattering models the term {\it model} is used advisedly.
The models in the literature are intended only to represent the approximate 
magnitude of strong $WW$ scattering cross sections. They
assume leading partial waves ($J=0$ and/or $J=1$ depending on the channel) 
that tend to saturate but not violate
unitarity. They are not complete quantum field theories and in particular the
unitarization methods typically violate crossing symmetry. These deficiencies
do not effect the utility of the models for the purpose
for which they are intended, and they 
are not addressed by the U-gauge method presented here, which merely
allows more information to be extracted within the spirit and limitations of 
the models.

The following sections present the basic idea, illustrate it  
for on-shell \\ 
$W^+W^+ \rightarrow W^+W^+$ scattering, apply it to the collider process 
$qq \rightarrow qqW^+W^+$, and
discuss some of the implications.

\noindent {\it \underline {The Basic Idea} }

Consider strong elastic scattering, $W_L^+ W_L^+ \rightarrow W_L^+ W_L^+$. 
To leading order in the $SU(2)_L$ coupling constant $g$ we 
decompose the amplitude into gauge
sector and symmetry-breaking sector contributions, 
$$
{\cal M}_{\rm Total} = {\cal M}_{\rm Gauge} + {\cal M}_{\rm SB}.\eqno(4)
$$
The gauge sector contribution is the sum of the 4-point Yang-Mills contact
interaction diagram and the  photon and $Z$ boson, $t$- and $u$-channel 
exchange diagrams. Each diagram makes a 
contribution that grows like $E^4$ where $E$ is the $W$ boson
center of mass energy. Gauge symmetry ensures that the terms proportional to
$E^4$ cancel leaving an O($E^2$) contribution, given by
$$
{\cal M}_{\rm Gauge}= -g^2\left(4-{3 \over \rho}\right){E^2 \over M_W^2} + 
{\cal M'}_{\gamma} + {\cal M'}_Z
\eqno(5)
$$
where $\rho = M_W^2/{\rm cos}^2\theta_W M_Z^2$ with $\theta_W$ 
the weak interaction
mixing angle. ${\cal M'}_{\gamma}$ and ${\cal M'}_Z$ are the residual
contributions of the photon and $Z$ exchange diagrams, of zero'th order in
$E$. (There is no residual interaction from the contact diagram.) 
For instance, the residual photon exchange amplitude, which contains the 
forward and backward Coulomb singularities, has the simple form,
$$
{\cal M'}_{\gamma} = -8g^2 {\rm sin}^2\theta_W (\beta^2 E^4)
{\beta^2 + (2+\beta^2){\rm cos}^2\theta \over ut}.
\eqno(6)
$$
where $\beta$ and $\theta$ are the $W$ velocity and scattering angle 
in the $WW$ center of mass, and $u,t = -2E^2\beta^2(1 \pm {\rm cos}\theta)$.

The order $E^2$ term in equation (5) is the ``bad high energy behavior''
that would render massive nonabelian gauge 
theories unrenormalizable were it not cancelled by the 
Higgs mechanism. The O($E^2$) term is also precisely the 
low energy theorem amplitude\cite{mc-hg-mg},
$$
{\cal M}_{\rm LET} = -\left(4-{3 \over \rho}\right){s \over v^2} = 
{\cal M}_{\rm Gauge} + {\rm O}(g^2)
\eqno(7)
$$
where $M_W=gv/2$ and $s=4E^2$.
The argument is simple: if the symmetry 
breaking force is strong, the quanta of the
symmetry breaking sector are heavy, $M_{SB} \gg M_W$, and 
decouple in gauge boson scattering at low energy, 
${\cal M}_{SB} \ll {\cal M}_{\rm Gauge}$. Then 
the quadratic term in ${\cal M}_{\rm Gauge}$ dominates 
${\cal M}_{\rm Total}$ for $M_W^2 \ll E^2 \ll M_{SB}^2$, which establishes the
low energy theorem to order $g^2$ without using the ET.
More familiar R-gauge derivations use the ET and current algebra or an
effective Lagrangian\cite{mc-hg-mg} to obtain the same result
from Goldstone boson scattering.\footnote{ 
The validity of the ET to all orders in $g$ is most natural in Landau gauge
(see Kilgore\cite{etrefs}), also a natural choice 
since the $w$ Goldstone bosons are indeed massless in Landau gauge.}

Even in the U-gauge method the starting point for strong scattering models 
is the R-gauge Goldstone boson amplitude, since it is in the Goldstone
boson amplitude that the strong dynamics is simply expressed, without the
cancellations that complicate the gauge boson amplitudes. 
In general we consider a strong scattering model labeled
``X'' for the unphysical Goldstone bosons,
${\cal M}^{\rm X}_{\rm Goldstone}(w^+w^+ \rightarrow w^+w^+)$. 
Combining the equivalence theorem,
equation (1), with equations (4) and (7), we find the model dependent
contribution of the symmetry breaking sector in U-gauge,
$$
{\cal M}_{\rm SB}^{\rm X}(W_LW_L) = {\cal M}^{\rm X}_{\rm Goldstone}(ww) - 
{\cal M}_{\rm LET}.
\eqno(8)
$$
up to corrections ${\rm O}(g^2,{M_W \over E})$.
We have used the ET to obtain the transcription from the 
Goldstone boson amplitude ${\cal M}^{\rm X}_{\rm Goldstone}$ to the equivalent 
symmetry breaking sector amplitude ${\cal M}^{\rm X}_{\rm SB}$ 
for physical $W_LW_L$ gauge boson scattering. The 
${\rm O}(g^2,{M_W \over E})$ corrections are inherent in any treatment of 
strong $WW$ scattering. 
Finally the complete gauge boson scattering amplitude is 
$$
{\cal M}^{\rm X}(W_LW_L)= {\cal M}_{\rm Gauge}(W_LW_L) + 
{\cal M}_{\rm SB}^{\rm X}(W_LW_L).
\eqno(9)
$$

\noindent {\it \underline {On-shell $W^+W^+ \rightarrow W^+W^+$ scattering} }

We illustrate the method for on-shell $W^+W^+$ scattering, considering the 
heavy Higgs boson model with $m_H=1$ TeV
and the K-matrix strong scattering model. 

In the Higgs boson model we can compare the
cross section obtained from our U-gauge method to the incoherently combined 
(``EWA'') cross section and to
the exact tree-level cross section. The starting point is the Goldstone boson
amplitude, 
$$
{\cal M}^{\rm Higgs}_{\rm Goldstone}(w^+w^+) = 
- { m_H^2 \over v^2} {t \over t -  m_H^2}
 + (t \rightarrow u).
\eqno(10)
$$
Applying equation (8) with $\rho =1$ and $s\simeq -t-u$ we obtain the U-gauge
transcription, 
$$
{\cal M}_{\rm SB}^{\rm Higgs}(W_L^+W_L^+) = -{t \over v^2} {t \over t -  m_H^2}
 + (t \rightarrow u),
\eqno(11)
$$
which differs from the exact U-gauge Higgs exchange amplitude by terms of order 
O($M_W^2/s$). 

Figure 1 compares the differential angular 
cross sections at $\sqrt{s}=1$ TeV. The three lower curves represent 
the results obtained from the incoherent sum 
($|{\cal M}^{\rm Higgs}_{\rm Goldstone}(ww)|^2 + |{\cal M}^{\rm Gauge}|^2$)
(dashed line), 
the coherent sum\\
$|{\cal M}_{\rm SB}^{\rm Higgs}(W_LW_L) + 
{\cal M}^{\rm Gauge}|^2$ (solid line), and
the exact tree level cross section (dot-dashed line). 
At $\theta = \pi/2$, the incoherent
and coherent approximations differ from the exact tree result by 13 and 7\%
respectively. In the forward direction, cos$\theta = 0.9$, 
where the Coulomb singularity is important, the incoherent
approximation differs from the exact value by 46\% while the coherent
approximation agrees to better than 3\%.

The K-matrix model is an arbitrary unitarization of the low energy theorem for 
the $J=0$, $I=2$ partial wave,
$$
{\cal M}^{\rm K}_{\rm Goldstone}(w^+w^+) = - 32\pi {x-ix^2 \over 1+x^2}
\eqno(12)
$$
where $x=s/32\pi v^2$. Like most strong $W^+W^+$ scattering model amplitudes 
but unlike the Higgs boson amplitude, 
${\cal M}^{\rm K}_{\rm Goldstone}(w^+w^+)$ 
cannot be expressed as a sum of $t$ and $u$ channel terms. Applying equation 
(8), the contribution to the $W_LW_L$ U-gauge amplitude is 
$$
{\cal M}^{\rm K}_{\rm SB}(W_L^+W_L^+) =  32\pi x^2 {x+i \over 1+x^2}.
\eqno(13)
$$
The angular cross sections from ($|{\cal M}^{\rm K}_{\rm Goldstone}(ww)|^2 +
|{\cal M}^{\rm Gauge}|^2$) (dashed line) and 
$|{\cal M}_{\rm SB}^{\rm K}(W_LW_L) + 
{\cal M}^{\rm Gauge}|^2$ (solid line) are displayed 
in the upper set of curves in figure 1.
The two agree to within 5\% at $\theta = \pi/2$ but disagree by 34\% at 
cos$\theta = 0.9$ where the incoherent approximation omits the large
interference contribution.

For the discussion of $qq \rightarrow qqWW$ it will be convenient to express
equation (13) in terms of an effective $s$-channel scalar propagator,
which for $W^+W^+$ would have charge and weak isospin $Q=I=2$.
We assign a conventional coupling $g M_W H^{--}_{\rm EFF} W^+_{\mu} 
 W^{\mu +}$ to this
fictitious object. Working only to leading order in $M_W^2/s$ we define the 
effective propagator,
$$
P_{\rm EFF}(s) = {-i \over 32\pi v^2} {x+i \over 1+x^2},
\eqno(14)
$$
so that its $s$-channel exchange reproduces equation (13).

In general for any model X in which ${\cal M}^{\rm X}_{\rm Goldstone}(ww)$ is a
function of $s$ alone, we can define 
$$
P_{\rm EFF}(s) = -i {v^2 \over s^2}({\cal M}^{\rm X}_{\rm Goldstone}
                 - {\cal M}_{\rm LET})
\eqno(15)
$$
so that the $s$-channel exchange  reproduces equation (8).

Notice that ${\cal M}_{\rm LET}(w^+w^+)=-s/v^2$ contributes 
$-i/s$ to $P_{\rm EFF}$, corresponding 
to a massless, scalar ghost. The unphysical singularity is of no concern since
our discussion is manifestly intended only for large values of $s$. In fact the 
apparent $Q=2$ $s$-channel ghost is just an artifact of our choice of 
an effective
$s$-channel interaction --- it can be viewed as arising from the $t$ and
$u$-channel exchanges of a massless (or light, i.e., $m\ltap {\rm O}(M_W)$)
$Q=0$ Higgs scalar propagating with a physical
(i.e., non-ghost) sign.\footnote{For $W^+W^- \rightarrow ZZ$ 
the contribution of ${\cal M}_{\rm LET}$ to $P_{\rm EFF}(s)$ would correspond 
to a $Q=0$ massless scalar propagating with a physical sign, i.e., a massless
Higgs boson.}
Massless scalar exchange and subtraction of ${\cal M}_{\rm LET}$ are
just alternate ways of representing the underlying physics that cancels 
the ``bad'' high energy behavior of the massive Yang-Mills interactions.
An alternative description of our procedure for transcribing strong 
$W^+W^+$ scattering
models to U-gauge is to represent the U-gauge symmetry
breaking sector by a massless standard model Higgs boson {\it plus} a 
$W^+_{\mu,L}W^{+,\mu}_LW^-_{\nu,L}W^{-,\nu}_L$ contact interaction term 
given by ${\cal M}^{\rm X}_{\rm Goldstone}(w^+w^+)$.

\noindent {\it \underline {$qq \rightarrow qqW^+W^+$ scattering} }

The real utility of the U-gauge method is in the application to 
$qq \rightarrow qqWW$ scattering, where we avoid the 
EWA and recover information
about the final state that is 
lost in the EWA. We begin by again decomposing the
amplitude into gauge sector and symmetry breaking sector components as in
equation (4). Now instead of 5 Feynman diagrams contributing to ${\cal M}^{\rm
Gauge}$ there are $\sim {\rm O}(100)$. For the $W^+W^+$ channel these 
include the five diagrams with 
$WW$ scattering topology in addition to diagrams in which one or both
final state $W$'s are radiated directly from a quark line. These gauge sector
diagrams, including the five with $WW$ scattering topology, are all calculated
exactly so that the cancellations among them required by gauge invariance 
are exactly fulfilled. 

In the diagrams with $WW$ scattering topology the ``initial state'' $W$'s are
virtual, with space-like masses of order $-q^2 \simeq {\rm O}(M_W^2)$. 
For pure s-wave strong scattering models, 
such as the K-matrix model, we 
parameterize the contribution of the symmetry breaking sector by the 
effective $s$-channel propagator $P_{\rm EFF}(s)$, equation (15). We are then
extrapolating the symmetry breaking sector contribution to its on-shell value.
The error 
introduced by the extrapolation is $\sim {s \over v^2} {-q^2 \over s}
\sim {\rm O}\left({M_W^2 \over v^2}\right) \sim {\rm O}(g^2)$. 
Essentially the same extrapolation
underlies the EWA\cite{soper-kunzst} and a similar one underlies the ET. 
(In the ET we extrapolate from 
gauge dependent Goldstone boson masses to $M_W$.)

In the results presented below this
prescription is applied to all $W$ polarization modes. The effect of
the ${\cal M}^{\rm X}_{\rm Goldstone}$ contact interaction on the
$W_TW_T$ and $W_LW_T$ scattering amplitudes is of the order of the 
O$(g^2)$ corrections intrinsic 
to any strong scattering ansatz. That this and other 
O$(g^2)$ approximations introduced by our U-gauge ``transcription'' are under
control is verified by the comparisons given below of cross sections
obtained by the EWA and the transcription method. 

Figure 2 compares the EWA and transcribed cross sections at the
LHC for $qq \rightarrow qqW^+W^+$ (for all $W$ polarizations and neglecting 
quark masses) using the K-matrix model and the 1 TeV Higgs
boson model. To simulate an 
experimentally relevant cross section a rapidity cut
$|\eta_W| < 1.5$ has been imposed. Eight curves are shown. 
In each case the dashed line is the EWA and
the solid line is the U-gauge transcription. The two upper pairs are the
total cross sections, while the lower pairs are the ``signals'' defined by
subtracting the standard model cross section with $m_H=0$. The larger 
signal is for the K-matrix model. 

Figure 2 shows that the cross sections from the two methods agree well, to the
extent that the lines are not easily distinguished in some cases. This is as
expected since the rapidity cut excludes the Coulomb singularity which 
would have caused them to differ. 
The total cross sections agree to within a few
percent over the range shown, while the signals differ by
about 10\% at $M_{WW}=600$ GeV and then converge to within a few percent as 
$M_{WW}$ increases.

\noindent {\it \underline {Discussion} }

The EWA is useful and computationally efficient, but it provides no
information on the rapidities and transverse momenta of the final state
quark jets nor on the net transverse momentum of the $WW$ diboson.
Since the EWA sets $p_T(WW)=0$,
it distorts the transverse momentum distributions of the
individual $W$ bosons and their decay products. The error is small for 
$p_T(W) \gg M_W$ but not at moderate $p_T(W)$.\footnote
{In earlier work this problem was addressed 
by smearing the EWA cross section with an
empirical $p_T(WW)$ distribution derived from heavy Higgs boson
production --- see \cite{ptsmear}.} The correct $p_T(WW)$ spectrum of each
model is automatically provided by the U-gauge transcription method.

The EWA neglects the interference between 
the gauge sector and symmetry breaking
sector amplitudes, so that it can only be reliably applied when one is much
bigger than the other. Thus the EWA computation is valid if
the signal is much larger than the standard model 
$qq \rightarrow qqWW$ background, but not necessarily if signal and background
are comparable. 
Since $W^+W^+$ is detected in the
like-charge lepton final state $l^+\nu l^+\nu$ and 
since solid angle coverage is
incomplete at any high energy collider, 
the $W^+_LW^+_L$ signal interferes not only with the 
$W^+_LW^+_L$ background but also with $W^+_LW^+_T$ and $W^+_TW^+_T$.
All these interference effects are automatically
included when the strong scattering models are embedded directly in
the complete set of diagrams for $ff \rightarrow ffWW$.

The  most serious shortcoming of the EWA is the inability to provide 
the final state jet distributions 
needed to evaluate the efficiencies of jet tag and veto strategies. 
A veto on events containing moderate-to-high $p_T$ jets at central 
rapidity effectively suppresses $qq \rightarrow qqW^+W^+$ standard model 
backgrounds at little
cost to the signal.\cite{cjv} A tag on higher rapidity, lower $p_T$ 
jets may be necessary to suppress the unexpectedly large background 
to $W^+W^+ \rightarrow l^+\nu l^+\nu$ from
$W^+Z \rightarrow l^+\nu l^+l^-$ in which the 
negative lepton escapes detection.\cite{WZbkgd} 

Jet tag and veto efficiencies for strong $WW$ scattering 
have been estimated using the complete set of tree diagrams for
$qq \rightarrow qqWW$ in the 
heavy Higgs boson ($m_H=1$ TeV) standard model.\cite{ptsmear,cjv} 
However the diboson energy spectrum in strong scattering models is
quite different, especially for
colliders of sufficient energy to avoid phase space suppression at 
$M_{WW}>1$ TeV. The
jet rapidity and transverse momentum spectra and the 
tag and veto efficiencies then 
also differ appreciably between strong scattering models and the heavy
Higgs boson model. 
By transcribing the strong scattering models
directly into the U-gauge amplitude, we compute the jet
distributions directly from the complete set of $qq \rightarrow qqWW$ 
tree diagrams, just as is done for Higgs boson models. The $p_T$ and $\eta$
distributions of the jets then correctly reflect the differing 
$WW$ energy distributions of the various strong scattering models.
In future work we will compare the distributions of 
the heavy Higgs boson and strong scattering models.

\vskip .2in
\noindent Acknowledgements: I wish to thank M.K. Gaillard and K. Hikasa for
helpful discussions. This work was supported
by the Director, Office of Energy
Research, Office of High Energy and Nuclear Physics, Division of High
Energy Physics of the U.S. Department of Energy under Contracts
DE-AC03-76SF00098 and DE-AC02-76CHO3000.

\newpage

\begin {center}
{\bf Figure Captions}
\end {center}

\noindent {\bf Figure 1.} Differential angular cross sections for on-shell 
$W^+_L W^+_L$
scattering at $\sqrt{s}=1$ TeV. The lower three curves are for the
Higgs boson model with $m_H = 1$ TeV. The upper two curves are for the
K-matrix model. In each case the dashed curve is obtained from 
the incoherent combination of ${\cal M}_{\rm Goldstone}$ and ${\cal M}_{\rm
Gauge}$ and the solid line is from the coherent gauge boson transcription.
The dot-dashed curve is the exact tree-level cross section for the Higgs
boson model.

\noindent {\bf Figure 2.} Cross sections for $W^+W^+$ production via 
$qq \rightarrow qqW^+W^+$ at the LHC with
$|\eta_W|<1.5$ for the K-matrix and 1 TeV Higgs boson models, computed by
the EWA (dashed lines) and the U-gauge method (solid lines). The two upper pairs
of curves are total cross sections (all $W$ polarizations) while the two lower 
pairs  are signal cross sections (predominantly longitudinal polarization) 
defined by subtraction of the standard model cross section with $m_H=0$. 
For both signal and background the larger pair of curves is for the K-matrix.

\newpage
\begin{figure}
\epsfbox[72 72 300 600]{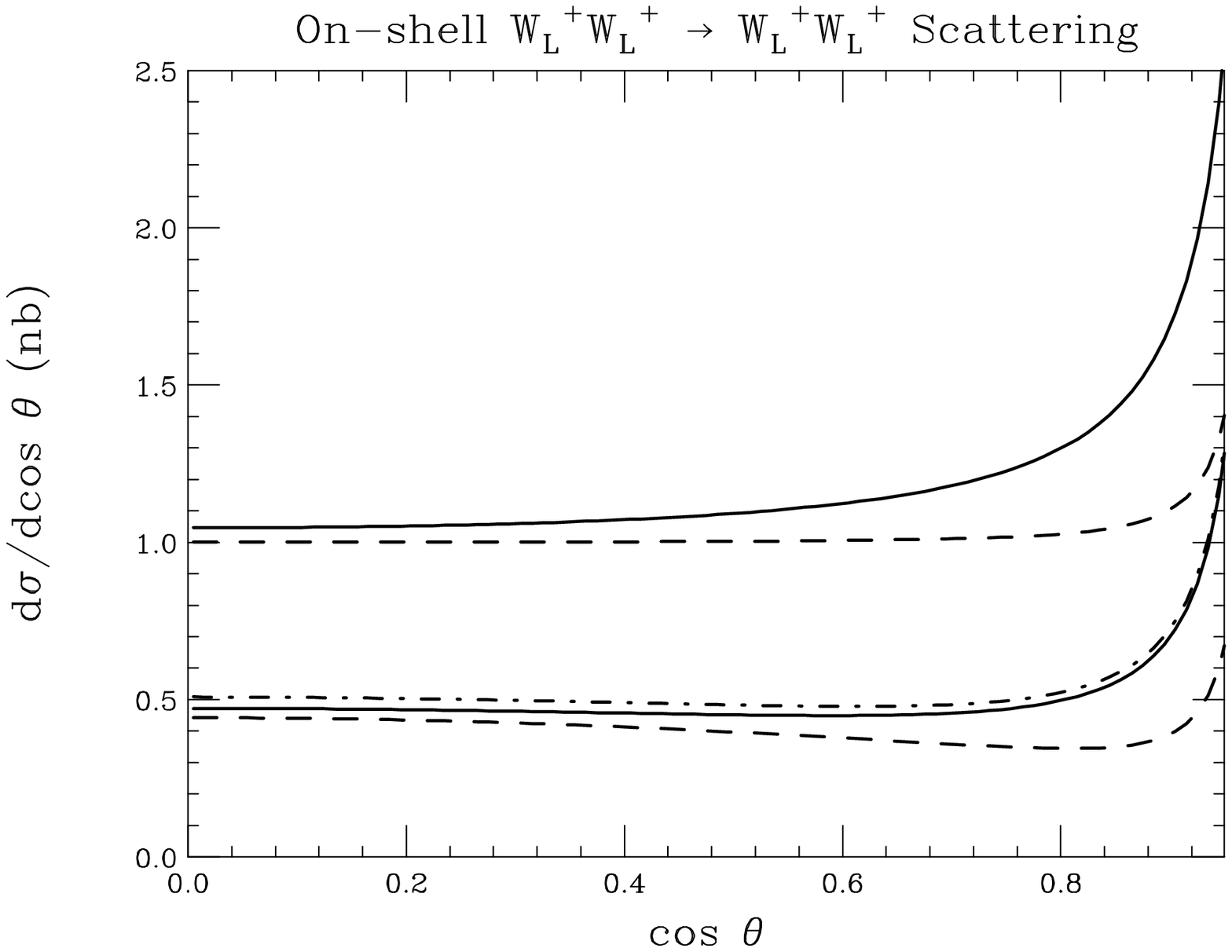}
%\label{fig1}
\end{figure}

\newpage
\begin{figure}
\epsfbox[72 72 300 600]{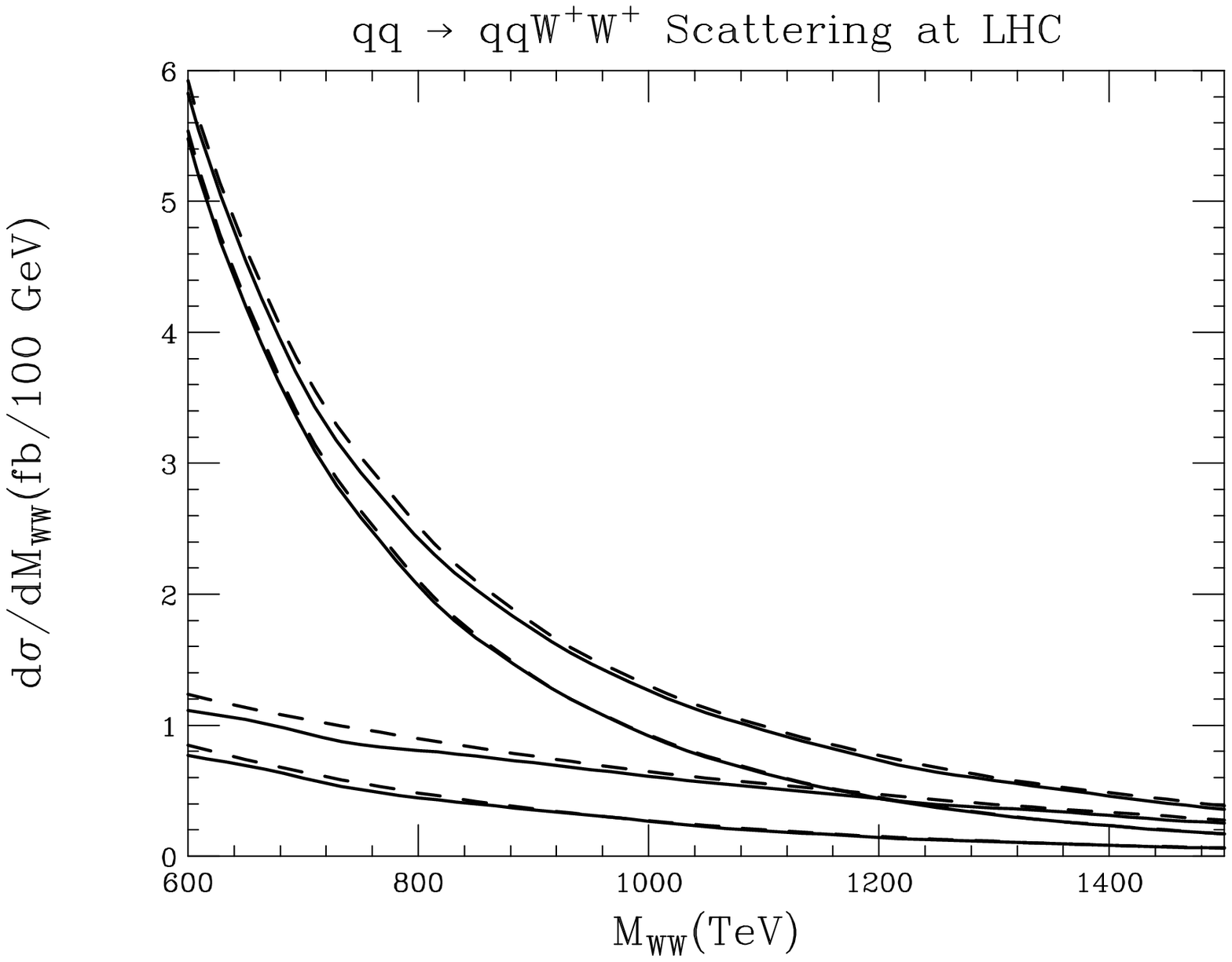}
%\label{fig2}
\end{figure}

\end {document}